\documentclass[
 reprint,
 amsmath,amssymb,
 aps,
]{revtex4-2}

\usepackage{graphicx}
\usepackage{dcolumn}
\usepackage{bm}
\usepackage{color}

\begin{document}

\preprint{APS/123-QED}

\title{Mean-field approximations of networks of spiking neurons with short-term synaptic plasticity}

\author{Richard Gast}%
 \email{rgast@cbs.mpg.de}
\author{Thomas R. Knösche}
 \altaffiliation[Also at ]{Institute for Biomedical Engineering and Informatics, TU Ilmenau, Germany}
\author{Helmut Schmidt}%
\affiliation{ 
Max Planck Institute for Human Cognitive and Brain Sciences, Leipzig, Germany
}%




\date{\today}

\begin{abstract}
Low-dimensional descriptions of spiking neural network dynamics are an effective tool for bridging different scales of organization of brain structure and function.
Recent advances in deriving mean-field descriptions for networks of coupled oscillators have sparked the development of a new generation of neural mass models.
Of notable interest are mean-field descriptions of all-to-all coupled quadratic integrate-and-fire (QIF) neurons, which have already seen numerous extensions and applications.
These extensions include different forms of short-term adaptation (STA) considered to play an important role in generating and sustaining dynamic regimes of interest in the brain.
It is an open question, however, whether the incorporation of pre-synaptic forms of synaptic plasticity driven by single neuron activity would still permit the derivation of mean-field equations using the same method.  
Here, we discuss this problem using an established model of short-term synaptic plasticity at the single neuron level, for which we present two different approaches for the derivation of the mean-field equations.
We compare these models with a recently proposed mean-field approximation that assumes stochastic spike timings.
In general, the latter fails to accurately reproduce the macroscopic activity in networks of deterministic QIF neurons with distributed parameters.
We show that the mean-field models we propose provide a more accurate description of the network dynamics, although they are mathematically more involved.
Using bifurcation analysis, we find that QIF networks with pre-synaptic short-term plasticity can express regimes of periodic bursting activity as well as bi-stable regimes. 
Together, we provide novel insight into the macroscopic effects of short-term synaptic plasticity in spiking neural networks, as well as two different mean-field descriptions for future investigations of such networks.
\end{abstract}

\maketitle


\section{\label{sec:intro}Low-Dimensional Manifolds of Spiking Neural Network Activity}

The brain can generate a variety of highly complex and chaotic patterns of neural activity \cite{basar_chaos_2012}.
However, given the vast number of neurons in the brain, these patterns appear to be less complex than they could be theoretically, indicating a high level of neuronal redundancy \cite{chialvo_emergent_2010,deco_emerging_2011}.
Electrophysiological recordings of macroscopic neural activity have revealed highly stereotyped responses to sensory stimulation as well as strongly synchronized regimes of neural activity \cite{engel_dynamic_2001,knosche_perception_2005,kujala_mismatch_2007,jirsa_nature_2014}.
More recently, multi-unit recordings have demonstrated that strong redundancies are present at the level of spiking neurons as well \cite{sadtler_neural_2014,murray_stable_2017}.
These findings indicate the existence of low-dimensional manifolds in the state space of the brain that typically govern its neural dynamics and its response to extrinsic stimulation.
The identification and description of such low-dimensional manifolds has been a central topic of neuroscientific research for many years \cite{babloyantz_low-dimensional_1986,kelso_dynamic_1995,celletti_low-dimensional_1996,bollimunta_neuronal_2008,spiegler_modeling_2011,deco_ongoing_2012}.
Different approaches for the derivation of mathematical descriptions of the temporal evolution of low-dimensional neural activity have been proposed \cite{deco_dynamic_2008}.
Among those are classic neural mass models that use direct, phenomenological descriptions of macroscopic measures of neural dynamics \cite{wilson_excitatory_1972,lopes_da_silva_model_1974,freeman_models_1978,jansen_electroencephalogram_1995,robinson_propagation_1997}.
For these neural mass models, equivalent spiking neural networks do not exist in general.  
Other approaches make use of probabilistic descriptions of the evolution of the collective behavior inside a neural population \cite{el_boustani_master_2009,buice_systematic_2009,schwalger_towards_2017}, which make it possible to capture the statistics inside the spiking neural network up to a certain order.
However, some of these approaches are restricted to asynchronous regimes of neural activity \cite{el_boustani_master_2009,buice_systematic_2009}, whereas others use approximations of random fluctuations in the spiking neural network \cite{schwalger_towards_2017}.  
Hence, neither of these approaches provide a mathematically exact set of mean-field equations that can describe the macroscopic dynamics of a spiking neural network in general.

The Ott-Antonsen ansatz has provided a new tool to derive mean-field models of spiking neural networks \cite{ott_low_2008}.
While originally devised for networks of all-to-all coupled Kuramoto oscillators \cite{kuramoto_collective_1991}, it has since been applied to networks of theta neurons \cite{luke_complete_2013,coombes_next_2019}, and, most relevant to this study, to networks of all-to-all coupled quadratic integrate-and-fire (QIF) neurons \cite{montbrio_macroscopic_2015}. 
For future applications of this method, it is of interest to know how well the derivation of the mean-field equations generalizes to other descriptions of neural dynamics than the particular QIF networks considered in \cite{montbrio_macroscopic_2015}.
Consequently, different extensions of the QIF model have been proposed that added biophysical mechanisms or structural details to the model in order to explain interesting neurodynamic phenomena, such as the onset of synchronized neural activity \cite{ratas_macroscopic_2016,byrne_mean_2017,di_volo_transition_2018,pietras_exact_2019,gast_mean-field_2020}.
Particularly interesting are extensions that include dynamic variables which are not driven by the mean-field activity of the network, but by neuron- or synapse-specific processes instead.
In such cases, it is unclear whether mean-field equations can still be found. 
In \cite{gast_mean-field_2020}, the QIF network was extended by a spike-frequency adaptation mechanism, where a neuron-specific adaptation current was elicited by the spiking activity of the same neuron.
Thus, the adaptation variable was not simply driven by the mean-field activity of the network.
To derive the mean-field equations nonetheless, the authors applied an adiabatic approximation to the adaptation dynamics.
This approximation assumes that the adaptation variable evolves slowly in comparison to the membrane potential dynamics and permits one to apply the mean-field derivation on the fast time-scale.
Based on this mean-field model it will be possible to investigate the effects of neuron-specific currents at meso- and macroscopic scales, such as for example the effects of calcium-dependent spikes on thalamic dynamics \cite{suffczynski_computational_2001} or the effects of spike-frequency adaptation on cortical microcircuits \cite{moran_neural_2007}.

In this work, we address the question of whether exact mean-field equations can be derived for QIF networks with synapse-specific dynamic variables.
Synaptic dynamics are especially interesting for the computational modeling of macroscopic neurodynamic phenomena.
This is because synaptic currents are thought to trigger the potential changes visible in macroscopic electrophysiological recordings of brain activity, and different synapse types come with different dynamic characteristics that are pivotal for our understanding of brain dynamics.
Classic neural mass models, for example, typically use different synaptic time scales to model rhythm generation in the brain \cite{lopes_da_silva_model_1974,jansen_electroencephalogram_1995,robinson_propagation_1997}.  
The QIF mean-field reduction generalizes to any convolution of the synaptic input with a synaptic response kernel \cite{montbrio_macroscopic_2015,ratas_macroscopic_2016} and, hence, allows one to derive mean-field descriptions of QIF networks with standard descriptions of synaptic dynamics such as the alpha kernel convolution \cite{jansen_electroencephalogram_1995,robinson_propagation_1997}. 
However, given appropriate stimulation, synaptic dynamics also undergo short-term plasticity (STP) that changes properties of the synaptic response.   
It has been shown that synapses can express short-term depression and facilitation and that time scales and strengths of these two STP forms differ between synapse and neuron types.
Moreover, synaptic STP has been linked to various functions and dynamic properties of the brain, such as working memory \cite{taher_exact_2020} or operating in a critical regime \cite{levina_dynamical_2007}. 
A generalization of the above discussed mean-field approaches to neural networks with synaptic STP would thus provide a valuable tool for modeling brain dynamics and function at the meso- and macroscopic level.

\begin{figure}[!ht]
    \centering
     \includegraphics[width=0.48\textwidth]{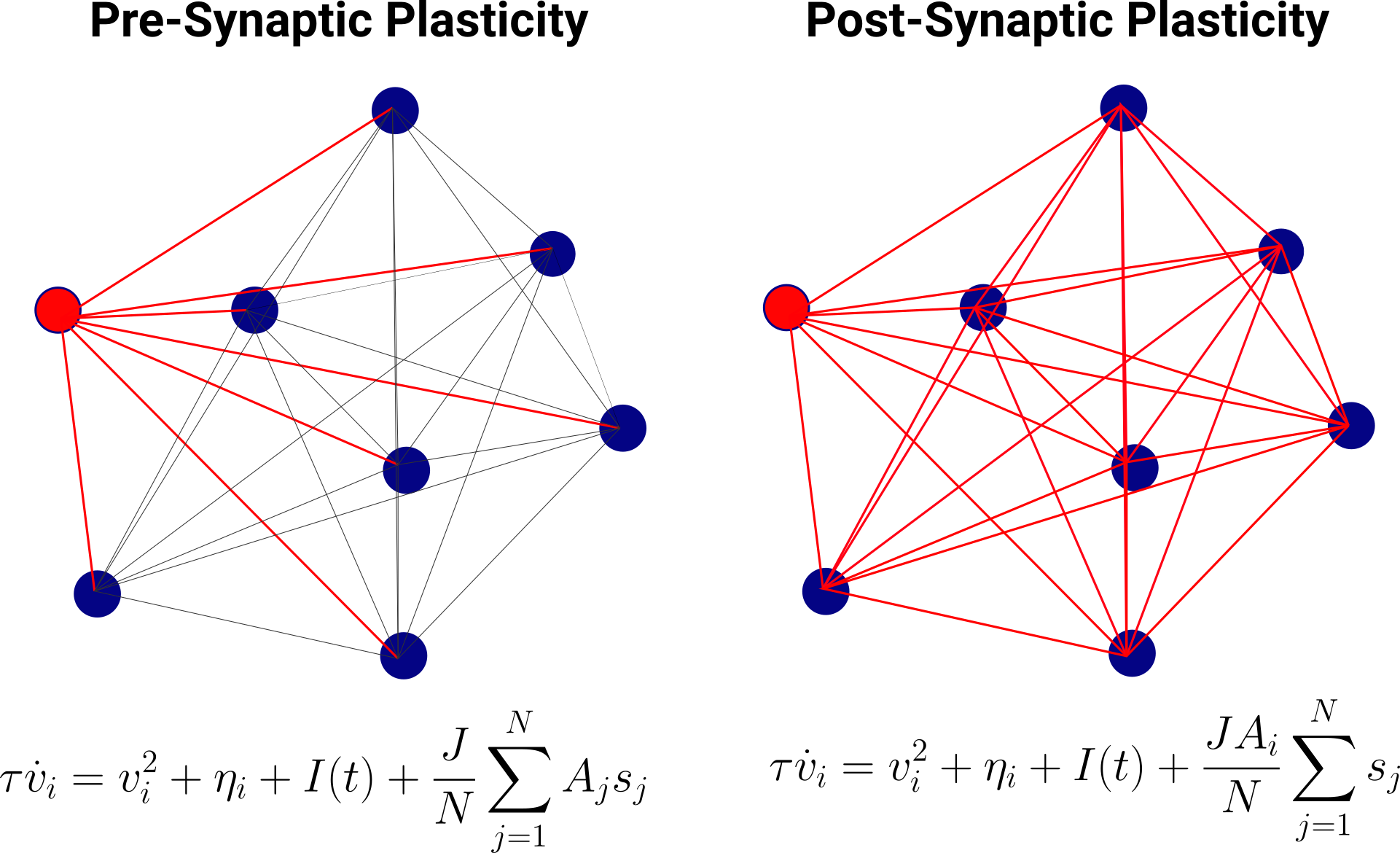}
    \caption{Pre- vs. Post-Synaptic Forms of Short-Term Plasticity. Nodes represent neurons in an all-to-all coupled network and edges between the nodes represent bidirectional synaptic couplings. Red nodes are active, i.e. did just spike, whereas blue nodes have not spiked for a sufficient period in time. Edges that are colored in red show adaptation in response to the activity of the red nodes, whereas grey edges do not. The two equations describe the membrane potential evolution of a QIF neuron for the cases of pre- and post-synaptic plasticity. Note that the adaptation variable $A_i$ is specific for pre-synaptic source neurons for the former case, and specific to post-synaptic target neurons for the latter.}
    \label{fig:synaptic_plasticity}
\end{figure}

Here, we discuss the descriptions of synaptic STP that are allowed for in the context of deriving Ott-Antonsen manifolds for heterogeneous QIF networks.
Recent work has demonstrated that mean-field equations can be derived for QIF networks with synaptic STP if two conditions are satisfied \cite{gast_mean-field_2020}: 
First, each time a neuron spikes in the network, it triggers synaptic STP at every other neuron, which is the case in all-to-all coupled networks.
Second, a single incoming spike triggers synaptic STP at all synapses of a neuron. 
Under those conditions, synaptic STP is no longer neuron specific and can simply be treated as a macroscopic variable driven by the mean-field activity of the network. 
This form of synaptic STP could be used to model forms of post-synaptic receptor desensitization, short-term changes in the number of available post-synaptic receptors, or resource depletion at the post-synaptic complex.
Importantly, it cannot be considered to represent pre-synaptic forms of plasticity, such as vesicle depletion.
While the first assumption would still hold for pre-synaptic STP in all-to-all coupled QIF networks, the second assumption would not.
Pre-synaptic resource depletion cannot be assumed to affect all network connections, but only the efferent connections of a specific neuron (see Fig.~\ref{fig:synaptic_plasticity}).

A well established model of pre-synaptic STP is the phenomenological model introduced in \cite{tsodyks_neural_1998}, which describes the dynamics of pre-synaptic facilitation and depression.
We will discuss the derivation of mean-field equations for QIF networks with pre-synaptic STP with respect to this model, though we will discuss the implications of our findings for general descriptions of pre-synaptic STP dynamics as well. 
In the following section, we define the microscopic model under consideration.
This will be followed by sections in which we discuss different approaches to derive equations for the low-dimensional network dynamics.
While we do not find the exact mean-field equations for QIF networks with pre-synaptic STP, we provide two different approximations that match well with the QIF network dynamics.
We point to the problems that would have to be solved in future attempts at an exact mean-field derivation and evaluate the accuracy of our approximate solutions via numerical simulations and bifurcation analysis.

\section{\label{sec:model}Low-Dimensional Manifolds of QIF Networks with STP}

We consider a network of $N$ all-to-all coupled QIF neurons with pre-synaptic STP
\begin{subequations}
    \begin{align}
        \tau \dot V_i &= V_i^2 + \eta_i + I(t) + \frac{J \tau}{N} \sum_{j=1}^N X_j^- U_j^+ S_j, \label{eq:vi} \\
        \tau_x \dot X_i &= 1 - X_i - \alpha X_i^- U_i^+ S_i \tau_x, \label{eq:xi}\\
        \tau_u \dot U_i &= U_0 - U_i + U_0(1 - U_i^-) S_i \tau_u, \label{eq:ui}\\
        S_i &=  \sum_{k \backslash t_i^k<t} \int_{-\infty}^t a(t-t') \delta(t'-t_i^k) dt', \label{eq:si}
    \end{align}
    \label{eq:qif_pre}
\end{subequations}
where eq.~\eqref{eq:si} represents a convolution of the spiking activity of neuron $i$ with a synaptic response kernel $a$, e.g. in the case of exponential synapses $a(t) = \mbox{e}^{-t/\tau_s}/\tau_s$ with synaptic time scale $\tau_s$.
A neuron $i$ emits its $k^{th}$ spike at time $t_i^k$ when it reaches a threshold $V_{\theta}$ upon which $V_i$ is reset to $V_r = -V_i$. 
Without loss of generality, we consider the limit $\tau_s \rightarrow 0$, such that $S_i$ represents the spiking activity of neuron $i$.
Eq.~\eqref{eq:xi} and eq.~\eqref{eq:ui} resemble the pre-synaptic STP mechanism described in \cite{tsodyks_neural_1998}.
We note here that $\cdot^-$ denotes a quantity just before a spike occurs (left limit), and $\cdot^+$ denotes a quantity just after the neuron spiked (right limit).
This discontinuity accounts for the biological fact that a pre-synaptic spike triggers synaptic facilitation before it can affect the post-synaptic neuron, by moving vesicles closer to the membrane.
Synaptic depression, however, results from the consumption of vesicles for the synaptic transmission process and is thus affected slightly later than synaptic facilitation.
We assume neural spiking activity to affect all outgoing synapses of a neuron equally, hence $X_i$ and $U_i$ can be considered as neuron- and not synapse-specific.
The adaptation dynamics are controlled by the depression and facilitation time constants $\tau_x$ and $\tau_u$, a depression strength $\alpha$, and a baseline synaptic efficacy $U_0$.
Eq.~\eqref{eq:vi} describes the evolution of the membrane potential $V_i$ of neuron $i$, which depends on a background excitability parameter $\eta_i$, an extrinsic forcing term $I(t)$, the membrane time constant $\tau$, and the coupling with the network activity.
The latter is given by a sum over the output $S_i$ of each neuron in the network, weighted by a global coupling strength $J$, and the neuron-specific synaptic depression $X_i$ and facilitation $U_i$.

In the limit $V_{\theta} \rightarrow \infty$, the membrane potential $V_i$ of a QIF neuron can be directly related to its phase via the transform $V_i = \tan(\frac{\theta_j}{2})$.
Under this transformation, (\ref{eq:vi}-\ref{eq:si}) represents a network of theta neurons \cite{ermentrout_parabolic_1986}, which can be considered a network of globally coupled oscillators.
Thus, the network satisfies the conditions for the existence of the Ott-Antonsen manifold, a low-dimensional manifold along which the network dynamics are guaranteed to evolve for $N \rightarrow \infty$ \cite{ott_low_2008,pietras_ott-antonsen_2016}.
This manifold can be described for (\ref{eq:vi}-\ref{eq:si}) by following the Lorentzian ansatz described in \cite{montbrio_macroscopic_2015}, i.e. by making the assumption that the state variables $V_i$ are distributed according to a Lorentzian where the probability density of $V$ for background excitability $\eta$ at time $t$ is given by 
\begin{equation}
    \rho(V|\eta,t) = \frac{1}{\pi} \frac{z(\eta,t)}{[V-y(\eta,t)]^2 + z(\eta,t)^2}.
    \label{eq:rho}
\end{equation}
The center $y(\eta, t)$ and half-width-at-half-maximum (HWHM) $z(\eta, t)$ of eq.~\eqref{eq:rho} are associated with the mean firing rate $r(\eta, t)$ and the membrane potential average over all neurons $v(\eta, t)$ via $z(\eta,t) = \pi r(\eta,t)$, and $y(\eta, t) = v(\eta, t)$, respectively.
Due to the conservation of the number of neurons, the network dynamics obey the following continuity equation:
\begin{equation}
    \partial_t \rho + \partial_V \left[ \left(\frac{V^2 + \eta + I}{\tau} + J r_{\mathrm{eff}} \right)\rho \right] = 0,
    \label{eq:cont}
\end{equation}
where $r_{\mathrm{eff}} = \frac{1}{N} \sum_{j=1}^N X_j^- U_j^+ S_j$ is the effective mean-field network activity that arrives at each neuron. 
By inserting eq.~\eqref{eq:rho} into eq.~\eqref{eq:cont} it can be shown that the dynamics of $z(\eta,t)$ and $y(\eta,t)$ obey
\begin{equation}
    \partial_t w(\eta,t) = i \left[ \frac{-w(\eta,t)^2 + \eta + I}{\tau} + J r_{\mathrm{eff}} \right],
    \label{eq:w}
\end{equation}
for any $\eta$, with $w(\eta,t) = z(\eta,t) + i y(\eta,t)$.
Without synaptic STP, i.e. for $U(t) = X(t) = 1$, eq.~\eqref{eq:w} can be solved for certain choices of the background excitability distribution.
The most drastic reduction in the dimensionality of the system can be achieved by choosing a Lorentzian distribution with density function
\begin{equation}
    g(\eta) = \frac{1}{\pi} \frac{\Delta}{(\eta - \bar \eta)^2 + \Delta^2},
\end{equation}
where $\bar\eta$ and $\Delta$ represent the center and HWHM of the distribution, respectively.
This choice allows one to solve
\begin{equation}
    \dot w = \int_{-\infty}^{\infty} \partial_t w(\eta,t) g(w) dw
\end{equation}
using the residue theorem of complex analysis, i.e. by evaluating the integral at the two poles of $g(w)$ given by $\bar\eta \pm i\Delta$.
Subsequently, eq.~\eqref{eq:w} can be solved for $r$ and $v$, yielding
\begin{subequations}
    \begin{align}
        \tau \dot r &= \frac{\Delta}{\pi\tau} +2 r v, \label{eq:r}\\
      \tau \dot v &= v^2 +\bar\eta + I(t) + J r \tau - (\pi r \tau)^2, \label{eq:v}
    \end{align}
\end{subequations}
where we additionally used $r_{\mathrm{eff}} = \frac{1}{N} \sum_{j=1}^N S_j = r$.

However, for non-constant $X$ and $U$, solving eq.~\eqref{eq:w} for $r$ and $v$ becomes a non-trivial problem.
In this case, $r_{\mathrm{eff}} = \frac{1}{N} \sum_{j=1}^N X_j^- U_j^+ S_j \neq r$ and, hence, $r_{\mathrm{eff}}$ must be calculated to arrive at closed-form equations for $r$ and $v$.
Two major problems have to be solved in this regard:
(a) The effective network input $r_{\mathrm{eff}}$ has to be expressed via mean-field variables such as the average firing rate $r$ and average depression and facilitation variables $x$ and $u$.
If this cannot be done, the mean-field equations would still contain neuron-specific variables, thus increasing their dimensionality dramatically. 
(b) The mean-field equations for the average depression $x = \frac{1}{N} \sum_{i=1}^N X_i$ and facilitation $u = \frac{1}{N} \sum_{i=1}^N U_i$ have to be solved.
However, the evaluation of these sums requires one to solve the coupled, non-linear differential equations \eqref{eq:xi} and \eqref{eq:ui}, which only has been achieved for stationary network input so far \cite{tsodyks_neural_1998}.
In the following section, we will address problem (b) and compare our results with recently proposed mean-field equations for a similar synaptic STP model \cite{schmutz_mesoscopic_2020}.
The remainder of this article will address different attempts to solve problem (a).

\section{\label{sec:stp_solution}Analytical solutions for microscopic STP}

As argued in the previous section, finding closed-form mean-field equations for the system given by equations~\eqref{eq:qif_pre} requires one to calculate the average depression $x = \frac{1}{N} \sum_{i=1}^N X_i$ and average facilitation $u = \frac{1}{N} \sum_{i=1}^N U_i$ across neurons.
We start by considering neuron $i$ that spikes periodically with a period $T$, thus producing a spike train $S_i(t) = \sum_{n=-\infty}^{\infty} \delta(t - nT_i)$.
The inter-spike interval $T_i$ corresponds to a firing rate of $1/T_i$.
In this scenario, solutions for the microscopic STP variables can be obtained analytically \cite{tsodyks_neural_1998}.
The evolution equations for synaptic short-term depression $X_i$ and short-term facilitation $U_i$ are given by eq.~\eqref{eq:xi} and eq.~\eqref{eq:ui}, respectively.
For the remainder of this section, we will omit the neuron index $i$ for brevity.
The (relative) strength of a synapse is given by $0 < U^+ X^- < 1$.
We denote $U$ by $U_n^-$ just before the corresponding neuron emitted its $n^{th}$ spike, and by $U_n^+$ just after the $n^{th}$ spike.
Solving the homogeneous part of the model equation, we obtain
\begin{equation}
    U_{n+1}^- = U_0 + (U_n^+ - U_0) \exp(-T/\tau_u),
\end{equation}
and the change of $U$ due to a spike is found to be
\begin{equation}
    U_{n+1}^+ = U_{n+1}^- + U_0 (1-U_{n+1}^-).
\end{equation}
These expressions can be reformulated into the following iteration scheme:
\begin{subequations}
    \begin{align}
      U_{n+1}^+ &= U_0 + (1-U_0) (U_0 + (U_n^+ - U_0) \mbox{e}^{-T/\tau_u}), \\
      U_{n+1}^- &= U_0 + (1-U_0) U_n^- \mbox{e}^{-T/\tau_u}.
    \end{align}
\end{subequations}
For the depression variable $X$, we find the following set of equations:
\begin{subequations}
    \begin{align}
      X_{n+1}^+ &= 1 + \left( (1 - \alpha U_n^+) X_n^- - 1 \right) \mbox{e}^{-T/\tau_x}, \\
      X_{n+1}^- &= (1-\alpha U_{n+1}^+) (1+(X_n^+ -1) \mbox{e}^{-T/\tau_x} ).
    \end{align}
\end{subequations}

In the stationary case, i.e. in the absence of transient dynamics, stationary solutions $U_\star^+ = U_n^+$, $U_\star^- = U_n^-$ and $X_\star^- = X_n^-, \, \forall n$ can be found:
\begin{subequations}
\begin{align}
  U_\star^+ &= \frac{U_0 + U_0(1-U_0)(1-\exp(-T/\tau_u))}{1-(1-U_0) \exp(-T/\tau_u)}, \\
  U_\star^- &= \frac{U_0}{1-(1-U_0) \exp(-T/\tau_u)}, \\
  X_\star^+ &= \frac{(1-\alpha U_\star^+) (1-\exp(-T/\tau_x))}{1-(1-\alpha U_\star^+)\exp(-T/\tau_x)}, \\
  X_\star^- &= \frac{1-\exp(-T/\tau_x)}{1-(1-\alpha U_\star^+)\exp(-T/\tau_x)}.
\end{align}
\label{eq:stat_xmin}
\end{subequations}
It is interesting to note that these results differ from the results when the firing rate is assumed to be a constant, i.e. when $S_i = r_0 = \rm{const.}$
In this case, we set $\dot U = \dot X = 0$, and obtain
\begin{subequations}
\begin{align}
  U_\star &= \displaystyle{\frac{U_0 + U_0 \tau_u r_0}{1 + U_0 \tau_u r_0}}, \\
  X_\star &= \displaystyle{\frac{1}{1 + \alpha \tau_x U^\star r_0}},
\end{align}
\label{eq:adapt_r0}
\end{subequations}
where we have made use of $U^+_\star = U^-_\star = U_\star$, as well as $X^+_\star = X^-_\star = X_\star$ since spike times are irrelevant.
The spike and rate description can be compared by equating $r_0 = 1/T$.

\begin{figure*}[!ht]
    \centering
    \includegraphics[width=1.0\textwidth]{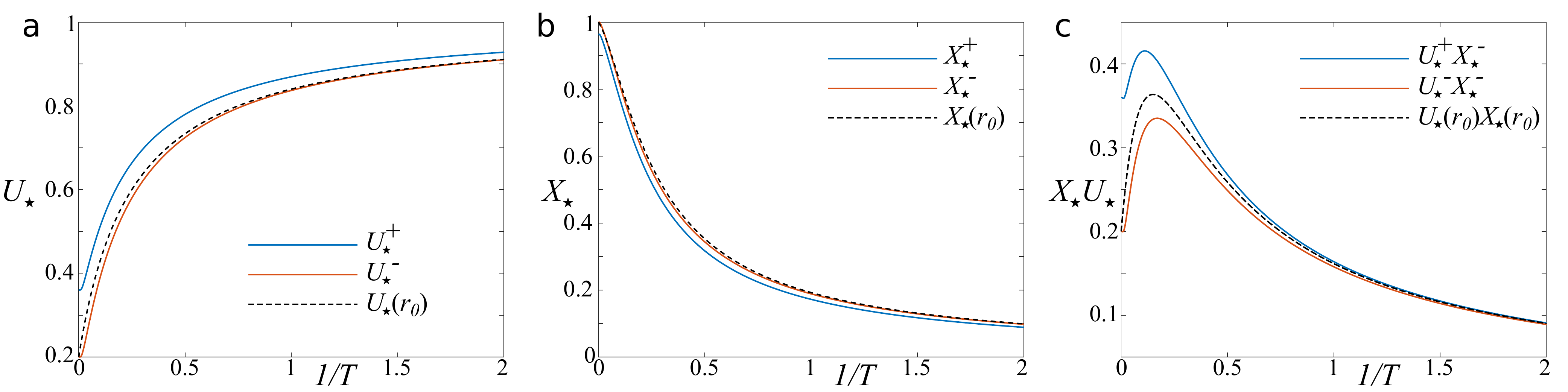}
    \caption{Comparison of the microscopic adaptation variables before and after spikes for discrete spikes, and for constant firing rates $r_0$. The inter-spike interval $T$ is varied. The constant firing rate is expressed as $r_0 = 1/T$. Parameters: $\alpha = 0.1$, $U_0 = 0.2$, $\tau_x = 50.0$, $\tau_u = 20.0$.}
    \label{fig:discrep1}
\end{figure*}

In figure~\ref{fig:discrep1} we compare these solutions for varying 
firing rates.
As can be seen, the results for constant firing rates $r_0$ are more closely related to the adaptation variables before spikes than after spikes.
This shows that it does matter for microscopic STP whether exact spike timings and the time of evaluation of $U$ and $X$ are considered or not, a finding which we expect to hold for non-stationary firing rates $S(t)$ as well. 

The expressions derived above can be used to evaluate the mean-field quantities $x$ and $u$, if the spike times or firing rates of all neurons are known.
Alternatively, they can be used to evaluate $r_{\mathrm{eff}}$ directly.
In the following sections, we will address the problem of evaluating $r_{\mathrm{eff}}$ to derive the mean-field equations for equations \eqref{eq:qif_pre}.
We will derive two different mean-field models, for which the results of this section will be used to refine the mean-field descriptions of the pre-synaptic STP dynamics.
In this context, we will evaluate how eq.~\eqref{eq:stat_xmin} vs. eq.~\eqref{eq:adapt_r0} affect the mean-field dynamics of the QIF network.

\section{\label{sec:poisson}Mean-Field Derivation Under a Poissonian Assumption of Neural Dynamics}

Recently, an approach for the derivation of a mean-field model for the system defined by eqs. \eqref{eq:qif_pre} has been presented in \cite{taher_exact_2020}.
The authors used a mean-field approximation of macroscopic quantities $x$ and $u$, averaged over all neurons in the network, that has been proposed in \cite{schmutz_mesoscopic_2020}.
In this article, a mean-field approximation of the effective network input
\begin{equation}
    r_{\mathrm{eff}}(t) = \frac{1}{N} \sum_{j=1}^N U_j^- X_j^- s_j,\label{eq:r_eff2}
\end{equation}
is derived, where $X_j^-$ and $U_j^-$ are given by eq.~\eqref{eq:xi} and eq.~\eqref{eq:ui}, respectively, with the modification that $U_j^+$ is replaced by $U_j^-$.
Whereas the original STP model formulation described in \cite{tsodyks_neural_1998} uses $U_j^+X_j^-$ as the effective weight of a synapse at the time of an incoming spike, Schmutz et al. use $U_j^-X_j^-$ instead \cite{schmutz_mesoscopic_2020}.
As shown in Fig.~\ref{fig:discrep1}C, these two choices can lead to substantial differences of the synaptic weight for small input rates.
Since an effective synaptic weight of $U_j^-X_j^-$ is also used in \cite{taher_exact_2020}, we will discuss the validity of their mean-field description for both the spiking neural network given by eq.~\eqref{eq:qif_pre} and the spiking neural network considered in \cite{taher_exact_2020}.
Henceforth, we will refer to the former as $\mathrm{SNN}_{\mathrm{pre}}$  and to the latter as $\mathrm{SNN}_{\mathrm{pre}}$ II.
Under the assumption that all $S_i$ follow independent Poisson processes, the effective network input in $\mathrm{SNN}_{\mathrm{pre}}$ II is approximated by $r_{\mathrm{eff}} \approx u(t) x(t) r(t)$, where $r(t)$ is the average firing rate across neurons at time $t$.
As explained in \cite{schmutz_mesoscopic_2020}, this mean-field approximation rests on two assumptions:
(I) Synapse indices can be randomized, i.e. the spike times matter, but not the synapses at which those spikes occur.
(II) The average impact of a spike on $X_i$ and $U_i$, $\forall i$ can be approximated by sampling from Gaussian distributions around the current values of $x$ and $u$.
A first-order mean-field approximation is then given by
\begin{subequations}
    \begin{align}
        \tau_x \dot x &= 1 - x - \alpha \tau_x x u r, \label{eq:x}\\
        \tau_u \dot u &= U_0 - u + U_0 \tau_u (1 - u) r. \label{eq:u}
    \end{align}
    \label{eq:tsodyks_mf_poisson}
\end{subequations}
As can be seen from these equations, both $x$ and $u$ are driven by the average firing rate $r = \frac{1}{N} \sum_{j=1}^N S_j$ of the QIF network.
This allows to one to apply the Lorentzian ansatz in the same way as demonstrated for post-synaptic depression in \cite{gast_mean-field_2020}.
The dynamics of the complex variable $w(\eta, t)$ can be expressed as 
\begin{equation}
    \partial_t w(\eta,t) = i[ \frac{-w(\eta,t)^2 +\eta + I(t)}{\tau} + J x u r],
    \label{eq:w_poisson}
\end{equation}
and by evaluating eq.~\eqref{eq:w_poisson} at $\pi r(t) + iv(t) = w(\bar \eta - i\Delta, t)$ one finds that the dynamics of $r$ and $v$ follow:
\begin{subequations}
\begin{align}
    \tau \dot r &= \frac{\Delta}{\pi\tau} +2 r v, \label{eq:r_poisson}\\
    \tau \dot v &= v^2 +\bar\eta + I(t) + J x u r \tau - (\pi r \tau)^2. \label{eq:v_poisson}
\end{align}
\label{eq:fre_poisson}
\end{subequations}
We will refer to the set of mean-field equations given by \eqref{eq:tsodyks_mf_poisson} and \eqref{eq:fre_poisson} as $\mathrm{FRE}_{\mathrm{Poisson}}$ where $\mathrm{FRE}$ stands for firing rate equations.

It is important to notice, however, that $\mathrm{FRE}_{\mathrm{Poisson}}$ cannot be considered exact.
While assumption (I) holds for a network of independent, homogeneous Poisson neurons (hence called Poissonian assumption), it does not hold in general \cite{schmutz_mesoscopic_2020}.
Therefore, the mean-field derivation essentially approximates a heterogeneous network of deterministic QIF neurons by a homogeneous network of stochastic Poisson neurons.
Furthermore, the first-order approximation given by eq.~\eqref{eq:x} and eq.~\eqref{eq:u} ignores the non-linear interaction between $X_i$ and $U_i$ in eq.~\eqref{eq:xi}.
As shown in \cite{schmutz_mesoscopic_2020}, considering second order dynamics can improve the accuracy of the mean-field approximation, especially in the vicinity of transient inputs to the network. 
Adding second-order dynamics would involve sampling from a multivariate Gaussian distribution over $(x, u)$, however.
This means that the mean-field derivation could not be considered deterministic and, hence, also not exact anymore.

\begin{figure*}
    \centering
    \includegraphics[width=1.0\textwidth]{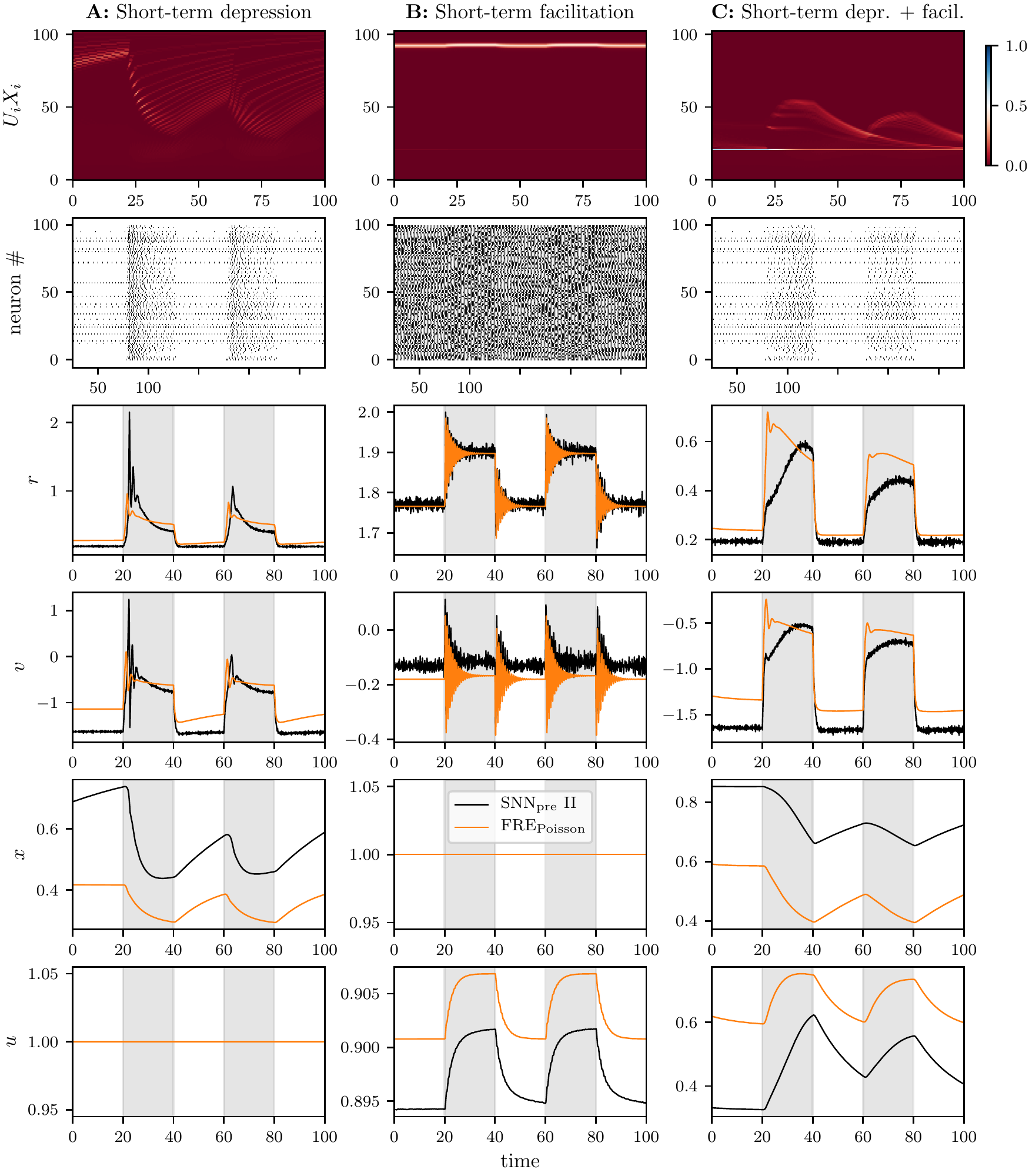}
    \caption{Evolution of the state variables of a QIF network and a mean-field approximation thereof for three different types of synaptic short-term plasticity (\textbf{A:} depression, \textbf{B:} facilitation, combined \textbf{C:} depression and facilitation). The first two rows show the distribution over the synaptic state $X_j U_j$ and the spiking activity of 100 randomly selected neurons, respectively. The last 4 rows show a comparison between the spiking neural network (black) and the mean-field approximation (orange) for the average firing rate $r$, the average membrane potential $v$, the average depression $x$, and the average facilitation $u$. In the SNN, averages were calculated across neurons $i$. Grey-shaded areas depict time intervals in which a rectangular input of $I(t) = 2.0$ was applied to the model. Color bars depict the probability density inside a given bin of the distribution over $X_i U_i$. Parameters for \textbf{A}: $U_0 = 1.0$, $\alpha = 0.1$. Parameters for \textbf{B}: $U_0 = 0.2$, $\alpha = 0.0$. Parameters for \textbf{C}: $U_0 = 0.2$, $\alpha = 0.1$. Other model parameters: $\tau = 1.0$, $\Delta = 2.0$, $\bar\eta = -3.0$, $J = 15.0\sqrt{\Delta}$, $\tau_x = 50.0$, $\tau_u = 20.0$, $N = 10000$.}
    \label{fig:micro_macro}
\end{figure*}

Still, it has been shown in \cite{taher_exact_2020} that $\mathrm{FRE}_{\mathrm{Poisson}}$ can accurately describe the mean-field dynamics of $\mathrm{SNN}_{\mathrm{pre}}$ II under certain conditions.
To test whether this holds in general, we compared the dynamics of the two models for three different STP parametrizations, leading to synapses that are either depressing, facilitating, or depressing and facilitating.
We solved the initial value problem of both sets of equations via an explicit Euler formalism with an integration step-size of $\mathrm{dt} = 0.0001$.
This step-size was sufficiently small to capture the dynamics of the network and was used for all subsequent numerical integration problems as well.
We then applied rectangular input pulses to the models and observed their dynamic responses around these inputs. 
The resulting time series can be observed in Fig.~\ref{fig:micro_macro}.
For purely depressing synapses, we find that there is a substantial mismatch between the mean-field dynamics of $\mathrm{SNN}_{\mathrm{pre}}$ II and $\mathrm{FRE}_{\mathrm{Poisson}}$.
As can be seen in Fig.~\ref{fig:micro_macro}A for the average depression $x$, there is a considerable offset between the mean-field model (orange) and the average of $X_i$ evaluated across neurons in the QIF network (black).
With respect to purely facilitating synapses, we find that the mean-field model provides a reasonable approximation of the QIF network.
Even though offsets can be observed between the mean-field model and the QIF network (see dynamics of $v$ in Fig.~\ref{fig:micro_macro}B), the qualitative behavior of the QIF network is captured well by the mean-field model.
This holds both in the steady-state regimes and during transient behavior around the on- and offsets of the input $I(t)$.
In the case of synapses with short-term depression and facilitation, the mean-field model expresses a substantial mismatch to the QIF network dynamics again.
For example, Fig.~\ref{fig:micro_macro}C shows that the dynamics of the average firing rate $r$ express focus dynamics for $\mathrm{FRE}_{\mathrm{Poisson}}$ after the onset of the first stimulus, whereas the average firing inside $\mathrm{SNN}_{\mathrm{pre}}$ II does not show such behavior.
In the upper row of Fig.~\ref{fig:micro_macro}, we show the evolution of the distribution over the combined synaptic state $X_i U_i$ in the microscopic model.
We find that this distribution tends to express multi-modalities in regions with a strong mismatch between mean-field and microscopic model.
These results suggest that the mean-field model can approximate the low-dimensional dynamics of the QIF network only if $X_i$ and $U_i$ express uni-modal, narrow distributions.
This finding makes intuitive sense, since the mean-field approximation of the dynamics of $U_i$ and $X_i$ given by eqs.~\eqref{eq:tsodyks_mf_poisson} represents a first order approximation.
Our results confirm that this approximation only performs well if the mean over $X_i$ and $U_i$ contains much information about the actual underlying distributions.
Thus, by providing these counter examples, we have shown that the mean-field model resulting from the Poisson assumption does not provide an exact mean-field description of the QIF network.

Since we are actually interested in the mean-field equations for $\mathrm{SNN}_{\mathrm{pre}}$ given by eqs.~\eqref{eq:qif_pre}, we now examine whether $\mathrm{FRE}_{\mathrm{Poisson}}$ can nonetheless provide an approximation of $\mathrm{SNN}_{\mathrm{pre}}$ under some conditions.
To gain further insight into the relationship between the mean-field equations and the QIF network, we asked whether there exists a QIF network description for which the mean-field model given by (\ref{eq:x}, \ref{eq:u}, \ref{eq:r_poisson}, \ref{eq:v_poisson}) can be considered exact.
Indeed, such a network exists and is easy to find.
Since $x$ and $u$ are only driven by the mean-field firing rate $r$, we can just introduce microscopic variables $U_i$ and $X_i$ that enter the microscopic evolution equation for $v_i$ in the same was as the macroscopic evolution equation for $v$ (\eqref{eq:v_poisson}) and are also driven by the mean-field activity of the QIF network:

\begin{subequations}
\begin{align}
    \tau \dot V_i &= V_i^2 + \eta_i + I(t) + \frac{J \tau}{N} U_i X_i s, \label{eq:vi_m} \\
    \tau_x \dot X_i &= 1 - X_i - \alpha X_i U_i s \tau_x, \label{eq:xi_m}\\
    \tau_u \dot U_i &= U_0 - U_i + U_0(1 - U_i) s \tau_u, \label{eq:ui_m}\\
    s &=  \sum_{j=1}^N \sum_{k \backslash t_j^k<t} \int_{-\infty}^t \delta(t'-t_j^k) dt', \label{eq:s_m}
\end{align}
\label{eq:qif_post}
\end{subequations}

where $s = r$ is the mean firing rate across all neurons in the network.
Apart from the description of the STP dynamics, this network description is equivalent to the one used in \cite{gast_mean-field_2020} for a QIF network with post-synaptic depression.
Indeed, under a first-order approximation of the dynamics of $x$ and $u$ via the Poissonian assumption, the system given by eqs.~\eqref{eq:qif_pre}, a QIF network with pre-synaptic STP, is essentially approximated by eqs.~\eqref{eq:qif_post}, a QIF network with post-synaptic STP (see Fig.~\ref{fig:synaptic_plasticity} for a visualization of the differences between the two).
Hence, we will refer to the network given by eqs.~\eqref{eq:qif_post} as $\mathrm{SNN}_{\mathrm{post}}$.

Next, we compared the behavior of the two different QIF network descriptions ($\mathrm{SNN}_{\mathrm{pre}}$ and $\mathrm{SNN}_{\mathrm{post}}$) to the mean-field model dynamics.
This was done to verify that $\mathrm{FRE}_{\mathrm{Poisson}}$ is indeed an exact mean-field model of $\mathrm{SNN}_{\mathrm{post}}$ and to see under which conditions pre- and post-synaptic STP have similar or different effects on the QIF network dynamics.
To this end, we used bifurcation analysis to identify phase transitions in the mean-field model around which we compared the behavior of the three models.
This way, we were able to set up stimulation paradigms that induce strong changes in the dynamic behavior of the mean-field model and evaluate whether the QIF networks express qualitatively similar phase transitions or not.
Bifurcation analysis was performed numerically, using the Python software PyRates \cite{gast_pyratespython_2019}, which provides an interface to the parameter continuation software Auto-07p \cite{doedel_auto-07p:_2007}.
We initialized the mean-field model with either purely depressing synapses ($U_0 = 1.0$, $\alpha = 0.04$) or purely facilitating synapses ($U_0 = 0.2$, $\alpha = 0.0$).
In each case, we performed a parameter continuation in the background excitability $\bar \eta$ for two different values of $\Delta \in {0.01, 0.4}$.
The latter introduces two different levels of firing rate heterogeneity to the QIF network.
We expected this firing rate heterogeneity to directly affect the broadness of the distributions over $X_i$ and $U_i$.
If that is indeed the case, the mean-field model should provide a better description of the $\mathrm{SNN}_{\mathrm{pre}}$ dynamics for $\Delta = 0.01$ than for $\Delta = 0.4$.

\begin{figure*}
    \centering
    \includegraphics[width=1.0\textwidth]{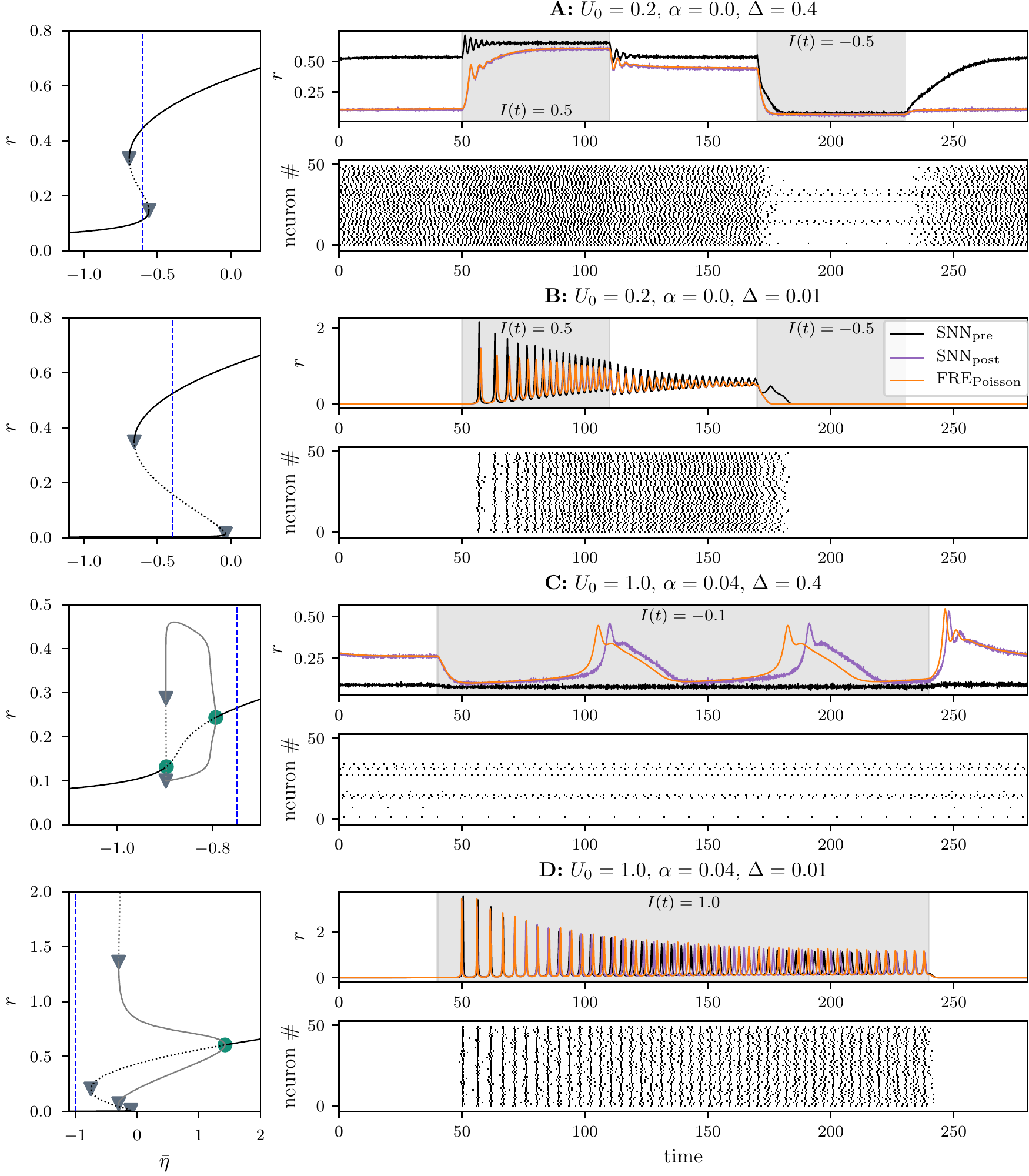}
    \caption{Comparison between $\mathrm{FRE}_{\mathrm{Poisson}}$ (orange), $\mathrm{SNN}_{\mathrm{pre}}$ (black), and $\mathrm{SNN}_{\mathrm{post}}$ (purple) for 4 different parameter sets (\textbf{A}-\textbf{D}). The first column shows 1D bifurcation diagrams in $\bar \eta$. Grey triangles represent fold bifurcations and green circles represent Andronov-Hopf bifurcations. Blue dashed lines mark the value of $\bar \eta$ that was used for the firing rate and spike raster plots in the second column. Spike raster plots show the spiking activity of 50 randomly selected neurons of $\mathrm{SNN}_{\mathrm{pre}}$. Grey shaded areas represent time intervals during which an extrinsic input $I(t)$ was applied to the models. Remaining model parameters: $J = 8.0$, $\tau_u = 20.0$, $\tau_x = 50.0$, $\tau = 1.0$, $N = 10000$}
    \label{fig:mf_bifurcations}
\end{figure*}

As can be seen in Fig.~\ref{fig:mf_bifurcations}A and B, we identified fold bifurcations for facilitating synapses for $\Delta = 0.4$ as well as $\Delta = 0.01$.
These fold bifurcations mark the outer limits of a bi-stable regime in which a stable high-activity focus and a stable low-activity node can co-exist, separated by a saddle-focus.
Indeed, we find that the steady-state behavior of the mean-field model and $\mathrm{SNN}_{\mathrm{post}}$ can be forced towards either of the two stable equilibria via extrinsic stimulation.
As shown for $\Delta = 0.4$ and $\Delta = 0.01$ in Fig.~\ref{fig:mf_bifurcations}A and B, respectively, there is always a very good agreement between those two models.
Regarding $\mathrm{SNN}_{\mathrm{pre}}$, we failed to identify the bi-stable regime for $\Delta = 0.4$.
In Fig.~\ref{fig:mf_bifurcations}A, it can be seen that the system behavior is only governed by a high-activity focus, even though the mean-field model predicts the co-existence of a low-activity stable node for $\bar \eta = -0.6$.
Thus, the mean-field model fails to predict the behavior of the QIF network with pre-synaptic STP in this case.
However, in the case of very low heterogeneity, we identified both stable states exists in $\mathrm{SNN}_{\mathrm{pre}}$ and found a good agreement with the mean-field model (see Fig.~\ref{fig:mf_bifurcations}B).

For depressing synapses, we found regimes of synchronized oscillations that emerge via Andronov-Hopf bifurcations for small as well as for high firing rate heterogeneity (see Fig.~\ref{fig:mf_bifurcations}C and D). 
Again, these oscillations could be induced in $\mathrm{FRE}_{\mathrm{Poisson}}$ as well as in $\mathrm{SNN}_{\mathrm{post}}$ with a very good match between the two.
Consistent with our findings for facilitating synapses, $\mathrm{SNN}_{\mathrm{pre}}$ expressed oscillations only for $\Delta = 0.01$ (see Fig.~\ref{fig:mf_bifurcations}D).
For higher firing rate heterogeneity ($\Delta = 0.4$), the network did not show any tendency to oscillate at all, even though the mean-field model predicted oscillations to be present at $\bar \eta = -0.85$ (see Fig.~\ref{fig:mf_bifurcations}C).

Thus, our results confirm that $\mathrm{FRE}_{\mathrm{Poisson}}$ is indeed an exact mean-field equation of $\mathrm{SNN}_{\mathrm{post}}$.
Furthermore, they demonstrate that $\mathrm{SNN}_{\mathrm{pre}}$ and $\mathrm{SNN}_{\mathrm{post}}$ can behave both very differently and very similarly, depending on the firing rate heterogeneity inside the network.
In our simulations, we were able to control this heterogeneity successfully via the parameter $\Delta$.
In regimes of low firing rate heterogeneity, $\mathrm{SNN}_{\mathrm{pre}}$ and $\mathrm{SNN}_{\mathrm{post}}$ expressed similar behavior, thus allowing for a good approximation of the mean-field dynamics of $\mathrm{SNN}_{\mathrm{pre}}$ via $\mathrm{FRE}_{\mathrm{Poisson}}$.
In regimes of high firing rates heterogeneity, the opposite was the case.
In the next sections, we investigate whether more accurate mean-field models of QIF networks with pre-synaptic STP can be derived and, if so, how they perform near the parameter regimes described in this section.

\section{Multi-population approximation of distributed parameters in the QIF network}

In the previous section, we have found that $\mathrm{FRE}_{\mathrm{Poisson}}$ is in good agreement with the dynamics of $\mathrm{SNN}_{\mathrm{pre}}$, when the distribution of $\eta_i$ is particularly narrow, i.e. when $\Delta \ll 1$.
Here, we exploit this fact and approximate the mean field dynamics by dividing the microscopic network into sub-networks with narrow distributions in $\eta_i$.
In other words, the Lorentzian distribution with $\{\bar \eta, \Delta\}$ is divided into a set of $M$ Lorentzian distributions with $\{\bar \eta_m, \Delta_m\}$, $m = 1,\ldots,M$, such that
\begin{equation}
    \frac{\Delta/\pi}{(\eta - \bar \eta)^2 + \Delta^2} \approx \frac{1}{M}\sum_{m=1}^M \frac{\Delta_m/\pi}{(\eta - \bar \eta_m)^2 + \Delta_m^2}.
\end{equation}

\begin{figure*}[!ht]
    \centering
    \includegraphics[width=1.0\textwidth]{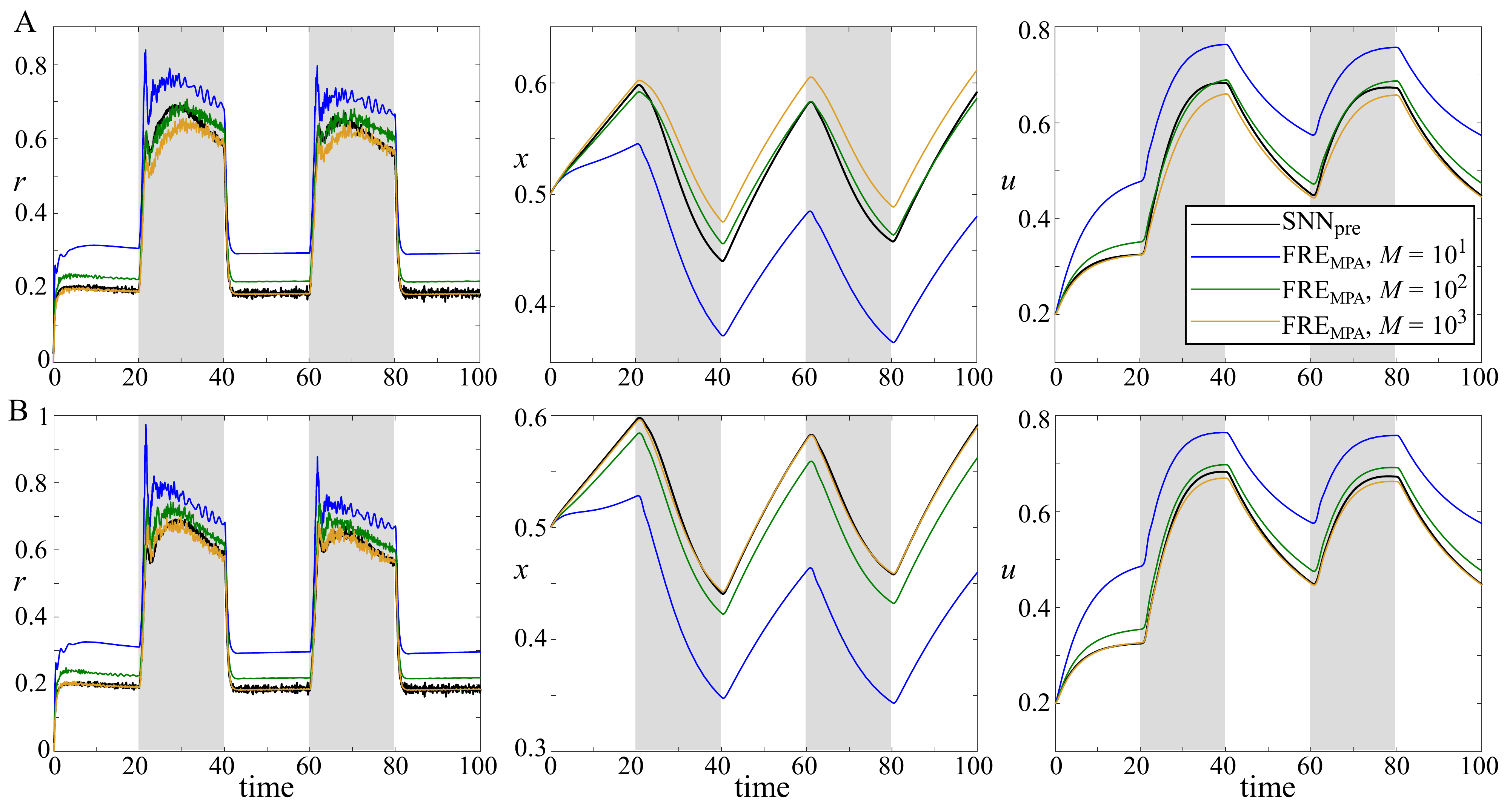}
    \caption{ Comparison of the mean field variables of the microscopic spiking neural network, and the mean field model of the spiking neural network divided into $M$ sub-networks with narrow distribution (multi-population approximation, MPA). Grey shaded areas indicate time intervals with $I(t) = 3.0$. {\bf A}: MPA with standard mean field description, {\bf B}: MPA with correction term for $U^+$. Parameters: $\alpha = 0.1$, $\tau = 1.0$, $\Delta = 2.0$, $\bar\eta = -3.0$, $J = 15.0 \sqrt{\Delta}$, $\tau_x = 50.0$, $\tau_u = 20.0$, $N = 10000$.}
    \label{fig:approx2}
\end{figure*}

The resulting set of equations for the evolution of the mean field variables is then given by
\begin{subequations}
\begin{align}
    \tau \dot r_m &= \frac{\Delta_m}{\pi\tau} +2 r_m v_m, \\
    \tau \dot v_m &= v_m^2 +\bar\eta_m + I(t) + \frac{J \tau}{M} \sum_{n=1}^M x_n u_n r_n - (\pi r_m \tau)^2, \\
    \dot x_m &= \frac{1-x_m}{\tau_x} - \alpha u_m x_m r_m, \\
    \dot u_m &= \frac{U_0 - u}{\tau_u} + U_0 (1-u_m) r_m.
\end{align}
\label{eq:approx2}
\end{subequations}
We will refer to this set of mean-field equations as $\mathrm{FRE}_{\mathrm{mpa}}$, for multi-population approximation.
One assumption we make here is that each sub-network contains the same number of neurons, which means that the weights for each sub-network are the same, and the mean field variables can be obtained by computing the mean $y = (1/M) \sum_{m=1}^M y_m$, where $y$ represents the mean field variable under consideration.
The parameters $\bar \eta_m$ and $\Delta_m$ are chosen as follows:
\begin{subequations}
\begin{alignat}{2}
    \bar \eta_m &= \mathrm{  } &\bar\eta + \Delta \tan \frac{\pi(2m-M-1)}{2(M+1)}, \\
    \Delta_m &= &\Delta (\tan \frac{\pi(2m-M-1/2)}{2(M+1)} \nonumber\\
    &  &- \tan \frac{\pi(2m-M-3/2)}{2(M+1)}).
\end{alignat}
\end{subequations}
The density of the parameters $\eta_m$ follows the Lorentzian distribution, and the $\Delta_m$ are chosen such that the half-widths approximately match the distances between the centers of the distributions of the sub-networks, i.e. $\bar \eta_{m+1} - \bar \eta_m \approx \Delta_{m+1} + \Delta_m$.
The results are shown in figure~\ref{fig:approx2}A. 
As can be seen, even at large $M$ the adaptation variables still show a small discrepancy with the result obtained from the spiking neural network $\mathrm{SNN}_{\mathrm{pre}}$.
We hypothesise that this difference is due to different results for the adaptation variables when the firing rate is assumed constant, and when it is assumed to be a spike train with constant ISI, as shown in Fig.~\ref{fig:discrep1}.
In other words, we expect that accounting for the fact that $\mathrm{FRE}_{\mathrm{Poisson}}$ was derived for $\mathrm{SNN}_{\mathrm{pre}}$ II instead of $\mathrm{SNN}_{\mathrm{pre}}$ will reduce the difference.
As the adaptation variables are in essence time-averaged quantities, the adaptation variables could be posed as $x = (X^- + X^+)/2$ and $u = (U^- + U^+)/2$.
However, with the update rules $U^+ = U^- + U_0 (1-U^-)$ and $X^+ = X^- - \alpha U^+ X^-$, this would yield out-of-bound values for $X^-$ at $x=1$, and $U^-$ at $u=0$.
The results shown in Figure~\ref{fig:discrep1} suggest that the mean field variables are closest to $X^-$ and $U^-$, which is why we set $X^- \approx x$, and $U^- \approx u$.
The update rule for $U^+$ gives the following correction term:
\begin{eqnarray}
    U^+(u) \approx u + U_0 (1-u).
\end{eqnarray}
Inserting this term into the mean field equations for $\mathrm{FRE}_{\mathrm{mpa}}$ produces a closer match of the mean field variables with the results of the microscopic model $\mathrm{SNN}_{\mathrm{pre}}$, see figure~\ref{fig:approx2}B.

\begin{figure*}[!ht]
    \centering
    \includegraphics[width=1.0\textwidth]{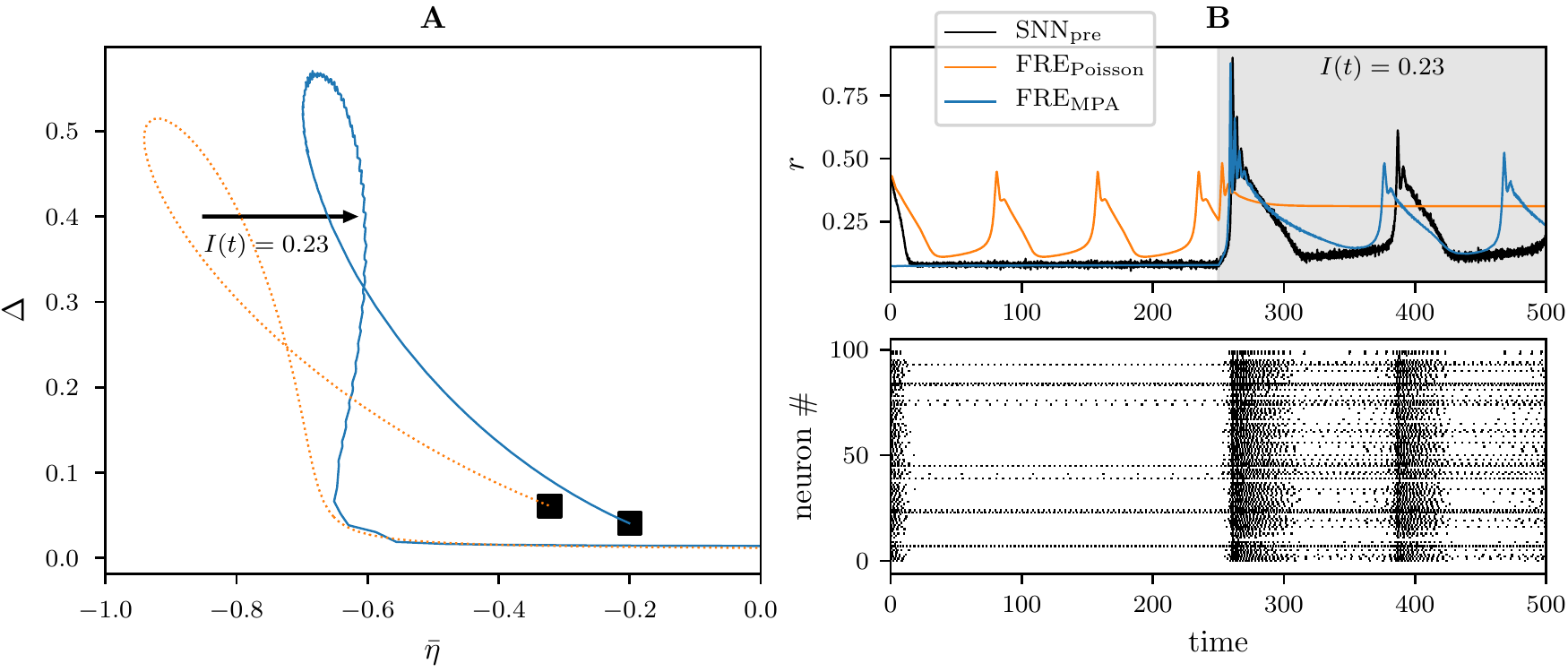}
    \caption{Phase transitions between steady-state and oscillatory regimes in $\mathrm{FRE}_{\mathrm{Poisson}}$ and $\mathrm{FRE}_{\mathrm{MPA}}$. \textbf{A:} 2D bifurcation diagram of the Hopf curve in $\mathrm{FRE}_{\mathrm{Poisson}}$ (orange) and $\mathrm{FRE}_{\mathrm{MPA}}$ (blue). The arrow represents the phase transition introduced by I(t) in either model. The black square represents the Bogdanov-Takens bifurcation from which the Hopf bifurcations emerge. \textbf{B:}The first row shows the simulated firing dynamics of the spiking neural network and both mean-field models. The second row shows the corresponding spiking activity of 100 randomly selected neurons of $\mathrm{SNN}_{\mathrm{pre}}$. Parameters: $\alpha = 0.04$, $U_0 = 1.0$, $\tau = 1.0$, $\Delta = = 0.4$, $\bar\eta = -0.85$, $J = 8.0$, $\tau_x = 50.0$, $\tau_u = 20.0$, $N = 10000$, $M = 100$, $I(t) = 0.23$ for $t > 250$ and $I(t) = 0.0$ otherwise.}
    \label{fig:mpa_bf}
\end{figure*}

As a final test of the predictive accuracy of $\mathrm{FRE}_{\mathrm{MPA}}$, we examined how well the model can predict the onset of oscillations in the QIF network.
Using bifurcation analysis, we identified the Hopf bifurcation leading to the oscillations in Fig.~\ref{fig:mf_bifurcations}C and investigated the locus of that Hopf bifurcation in the 2D parameter space spanned by $\bar\eta$ and $\Delta$.
This, we did for both $\mathrm{FRE}_{\mathrm{Poisson}}$ and $\mathrm{FRE}_{\mathrm{MPA}}$ with $M = 100$ mean-field populations.
As shown in Fig.~\ref{fig:mpa_bf}A, we found that the Hopf curves emerged from a Bogdanov-Takens bifurcation in both $\mathrm{FRE}$ models.
This represents the same bifurcation structure as has already been identified for QIF networks with SD (see Fig.2 and 4 in \cite{gast_mean-field_2020} for the corresponding 1D and 2D bifurcation diagrams, respectively).
Furthermore, we have shown the corresponding 1D bifurcation diagrams for the $\mathrm{FRE}_{\mathrm{Poisson}}$ model for $\Delta = 0.4$ and $\Delta = 0.01$ in Fig.~\ref{fig:mf_bifurcations}C and D, respectively.
Thus, we expect stable oscillations to exist in the regions enclosed by the Hopf curves.
As shown in Fig.\ref{fig:mpa_bf}A, the difference between the Hopf curves predicted by $\mathrm{FRE}_{\mathrm{Poisson}}$ and $\mathrm{FRE}_{\mathrm{MPA}}$ becomes larger when $\Delta$ increases. 
For $\Delta = 0.4$, $\mathrm{FRE}_{\mathrm{Poisson}}$ predicts stable oscillations to exist at $\bar\eta = -0.85$, which we already failed to find in the QIF network in Fig.\ref{fig:mf_bifurcations}D.
$\mathrm{FRE}_{\mathrm{MPA}}$ predicts the existence of a stable node at $\bar\eta = -0.85$, however, and the existence of stable oscillations for $-0.66 < \bar\eta < -0.6$.
To see whether the oscillations predicted by $\mathrm{FRE}_{\mathrm{MPA}}$ indeed exist in $\mathrm{SNN}_{\mathrm{pre}}$, we performed numerical simulations where we initialized the QIF network at $\bar\eta = -0.85$ and then forced it towards $\bar\eta = -0.62$ via extrinsic stimulation.
As can be seen in Fig.\ref{fig:mpa_bf}B, the QIF network expressed steady-state behavior for $\bar\eta = -0.85$ and started to oscillate when pushed to $\bar\eta = -0.62$.
Hence, $\mathrm{FRE}_{\mathrm{MPA}}$ correctly predicted the existence of oscillatory bursts in the QIF network for $ M = 100$, but not for $M = 1$, for which $\mathrm{FRE}_{\mathrm{MPA}}$ reduces to $\mathrm{FRE}_{\mathrm{Poisson}}$.
The bursts have similar properties as the ones found in QIF networks with post-synaptic plasticity \cite{gast_mean-field_2020} and can be expected to result from the interaction between synaptic short-term depression and recurrent excitation via the network.
Comparing the firing rate dynamics of $\mathrm{FRE}_{\mathrm{MPA}}$ and $\mathrm{SNN}_{\mathrm{pre}}$ in Fig.\ref{fig:mpa_bf} reveals a slight difference between the oscillation period of the mean-field model and the QIF network.
This difference shows that $\mathrm{FRE}_{\mathrm{MPA}}$ can not be considered an exact mean-field model, even for $M = 100$.
Still, we find that it captures the phase transitions inside $\mathrm{SNN}_{\mathrm{pre}}$ well and thus provides a reasonable trade-off between accuracy and computational complexity.

\section{\label{sec:slow}Adiabatic Approximation of STP Dynamics}

For simplification, we will consider synapses with mere short-term depression in this section, since we showed in section~\ref{sec:poisson} that the mismatch between the mean-field model $\mathrm{FRE}_{\mathrm{Poisson}}$ and the QIF networks $\mathrm{SNN}_{\mathrm{pre}}$ and $\mathrm{SNN}_{\mathrm{pre}}$ II could be reproduced in this simpler case as well.
We thus consider the microscopic system given by
\begin{subequations}
\begin{align}
    \tau \dot V_i &= V_i^2 + \eta_i + I(t) + \frac{J \tau}{N} \sum_{j=1}^N X_j^- S_j, \label{eq:vi_d} \\
    \tau_x \dot X_i &= 1 - X_i - \alpha X_i^- S_i \tau_x, \label{eq:xi_d}\\
    S_i &=  \sum_{k \backslash t_j^k<t} \int_{-\infty}^t a(t-t') \delta(t'-t_j^k) dt'. \label{eq:si_d}
\end{align}
\end{subequations}
In this system, we approximate the STP dynamics via a linear differential operator $L$, i.e. $L X_i(t) = S_i(t)$. 
In such a case, a Green's function $G(t)$ exists that allows one to express the dynamics of $X_i$ via a convolution of $G(t)$ with the spiking activity of neuron $i$:
\begin{equation}
   X_i(t) = \int_{-\infty}^t G(t-t') S_i(t') dt' = G * S_i.
\end{equation}
Then, since $S_i$ is related to $z(\eta_i, t)$ via $S_i \pi = z(\eta_i, t)$, eq.~\eqref{eq:w} can be written as
\begin{equation}
    \partial_t w(\eta,t) = i[ \frac{-w(\eta,t)^2 +\eta + I(t)}{\tau} + J (G * \frac{\Re[w]}{\pi})\Re[w]].
    \label{eq:w2}
\end{equation}
To solve eq.~\eqref{eq:w2} for $r$ and $v$, the effective firing rate $r_{\mathrm{eff}} = \int_{-\infty}^{\infty} (G * r(\eta)) r(\eta) g(\eta) \mbox{d}\eta$ must be determined, which requires one to evaluate the product between the single cell firing rate and a convolution of itself.
This makes it difficult to find a closed-form solution for $r$ and $v$, since the synaptic depression kernel $G$ cannot simply be pulled out from the convolution integral.
The simplest approximation of this problem is to replace the convolution integral by a mean synaptic depression, as is done for the Poissonian assumption.
Alternatively, we assume that the dynamics of $X_i$ are slow in comparison to the dynamics of $v_i$.
For the relaxation dynamics of $X_i$, this assumption is met if $\tau_x \gg \tau$.
We note here, however, that the spiking activity of the neuron also introduces a relatively fast time scale to eq.~\eqref{eq:xi_d}, which may violate our assumption.
Still, under this assumption, we can apply an adiabatic approximation to the system and consider the dynamics of the fast sub-system for effectively constant adaptation (see \cite{gigante_diverse_2007,gast_mean-field_2020} for a similar approach):
\begin{subequations}
\begin{align}
    \tau \dot V_i &= V_i^2 + \eta_i + I(t) + \frac{J \tau}{N} \sum_{j=1}^N X_j^- S_j, \label{eq:vi_fast} \\
    S_i &=  \sum_{k \backslash t_j^k<t} \int_{-\infty}^t \delta(t'-t_j^k) \mbox{d}t', \label{eq:si_fast}
\end{align}
\end{subequations}
where $X_j$ is approximated as neuron-specific constant.
Due to the Lorentzian distribution of the background excitabilities $\eta_i$ and the resulting heterogeneity of single cell firing rates in the network, $X_i$ cannot be assumed as homogeneous across neurons.
Instead, it must be considered a distributed quantity, governed by a probability density function $h(X_i)$.
Then, the main difficulty in developing the mean field description lies in the fact that $h(X_i)$ is generally unknown if a mean field variable is considered.
More precisely, if we consider the mean field variable $x$ that describes the average synaptic depression across the network, little is known about the distribution of the microscopic variables $X_i$, which is required to determine the effective firing rate $r_{\mathrm{eff}}$. 
By using the adiabatic approximation, we argue that an approximation of $r_{\mathrm{eff}}$ can be obtained by estimating the distributions $X(\eta)$ and $r(\eta)$ from the mean field variables in the stationary case, and solving
\begin{equation}
    r_{\mathrm{eff}} = \int_0^1 \int_{-\infty}^{\infty} X r(\eta) h(X|\eta) g(\eta) \mbox{d}\eta \mbox{d}X.\label{eq:r_eff}
\end{equation}
Assuming independent Lorentzian density functions for $h$ and $g$, i.e. $h(X|\eta)g(\eta) = h(X)g(\eta)$, eq.~\eqref{eq:w2} would only need to be evaluated at the poles in the lower half-planes $\pi r(t) + iv(t) = w(\bar\eta-i\Delta, \bar X - i\Delta_X, t)$, where $\bar X$ and $\Delta_X$ would represent the center and HWHM of the Lorentzian distribution over $X$, respectively.
Then, the effect of pre-synaptic STP on the network dynamics would effectively reduce to a distribution over the coupling parameter $J$.
For the mean-field equations of a QIF network with distributed coupling parameters see \cite{montbrio_macroscopic_2015}.
However, $h$ and $g$ cannot be assumed to be independent, since $\eta_i$ controls the firing rate of neuron $i$, which in turn controls its synaptic depression $X_i$.
Furthermore, $X$ is bound between $[0, 1]$ and hence a Lorentzian distribution cannot be assumed.
In the upper row of Fig.~\ref{fig:micro_macro}, we show the evolution of the distribution over $X_i U_i$ for three different parametrizations, corresponding to a purely depressing synapse, a purely facilitating synapse, and a synapse with facilitation and depression acting on different time scales.
Importantly, the evolution of the distribution reveals that it is not always uni-modal.
For purely depressing synapses, it clearly expresses an at least bi-modal distribution over the whole time course.
Thus, finding an appropriate form of $h$ that holds in general is a highly non-trivial problem that we did not find a solution for.

To further simplify the problem, we assume that the depression of a neuron's efferent synapses $X_i$ is merely a function of the firing rate $r_i$ of the same neuron.
The stationary firing rate of a QIF neuron in response to an external Input $I_{in}$ is $\sqrt{I_{in}}/\pi$ if $I_{in} > 0$, and zero otherwise.
Hence, the distribution of firing rates for a given input is (in the stationary case) given by
\begin{equation}
    r(\eta;I_{in}) = H(\eta+I_{in}) \sqrt{\eta + I_{in}}/\pi,
\end{equation}
where $H$ is the Heaviside step function.
Therefore, for any given mean field firing rate $r$ one can find a unique constant $I_r$ for which 
\begin{equation}
    r = \int_{-\infty}^{\infty} r(\eta;I_r) g(\eta) \mbox{d}\eta,
\end{equation}
which allows us to translate the mean field variable $r$ into the distribution $r(\eta;I_r)$.

Similarly, we can use the assumption that $X_i$ is a function of $r_i$ to translate the mean field variable for synaptic depression, $x$, into the distribution $X(\eta;I_x)$.
First, we use the rate relationship given by eq.~\eqref{eq:adapt_r0} to approximate
\begin{equation}
    x(\eta;I_x) = 1/(1+\alpha \tau_x r(\eta;I_x)),
    \label{eq:xeff_rate}
\end{equation}
for any given input $I_x$, and then define
\begin{equation}
    x_1 = \int_{-\infty}^{\infty} \rho(\eta)/(1+\alpha \tau_x r(\eta;I_x)) \mbox{d}\eta.
\end{equation}
Alternatively, we can use eq.~\eqref{eq:stat_xmin} to approximate the distribution $x(\eta)$ in the spiking scenario:
\begin{equation}
    x(\eta;I_x) = \frac{1-\exp(-1/\tau_x r(\eta;I_x))}{1-(1-\alpha)\exp(-1/\tau_x r(\eta;I_x))},
    \label{eq:xeff_spike}
\end{equation}
which yields
\begin{equation}
    x_2 = \int_{-\infty}^{\infty} \frac{(1-\exp(-1/\tau_x r(\eta;I_x))) g(\eta)}{1-(1-\alpha)\exp(-1/\tau_x r(\eta;I_x))} \mbox{d}\eta.
\end{equation}

Having obtained $I_r$ and $I_x$, we can ultimately compute
\begin{equation}
    r_{\mathrm{eff}} = \int_{-\infty}^{\infty} r(\eta;I_r) x(\eta;I_x)) g(\eta) \mbox{d}\eta,
\end{equation}
where $x(\eta;I_x)$ is either chosen for the rate scenario (eq.~\eqref{eq:xeff_rate}), or in the spike scenario (eq.~\eqref{eq:xeff_spike}).
This requires one to solve
\begin{equation}
    r_{\mathrm{eff}} = \frac{\Delta}{\pi^2}\!\!\!\!\!\!\!\!\!\! \int\displaylimits_{\mathrm{min}(-I_x,-I_r)}^{\infty} \!\!\!\!\!\!\!\!\!\! \frac{1}{1\!+\!\alpha \tau_x \sqrt{\eta\!+\!I_x}} \frac{\sqrt{\eta+I_r}}{(\eta-\bar\eta)^2 + \Delta^2} \mathrm{d} \eta,
\end{equation}
in the rate scenario, and
\begin{equation}
    r_{\mathrm{eff}} = \frac{\Delta}{\pi^2}\!\!\!\!\!\!\!\!\!\! \int\displaylimits_{\mathrm{min}(-I_x,-I_r)}^{\infty} \!\!\!\!\!\!\!\!\!\! \frac{\exp\left(\frac{\pi}{\tau_x \sqrt{\eta+I_x}}\right)-1}{\exp\left(\frac{\pi}{\tau_x \sqrt{\eta+I_x}}\right)\!-\!(1\!-\!\alpha)} \frac{\sqrt{\eta+I_r}}{(\eta-\bar\eta)^2 + \Delta^2} \mathrm{d} \eta,
\end{equation}
in the spiking scenario.
We refer to this mean-field model as $\mathrm{FRE}_{\mathrm{aa}}$ for adiabatic approximation, with $\mathrm{FRE}_{\mathrm{aa1}}$ and $\mathrm{FRE}_{\mathrm{aa2}}$ denoting the mean-field model considering the rate and spike scenario, respectively.

\begin{figure*}[!ht]
    \centering
    \includegraphics[width=1.0\textwidth]{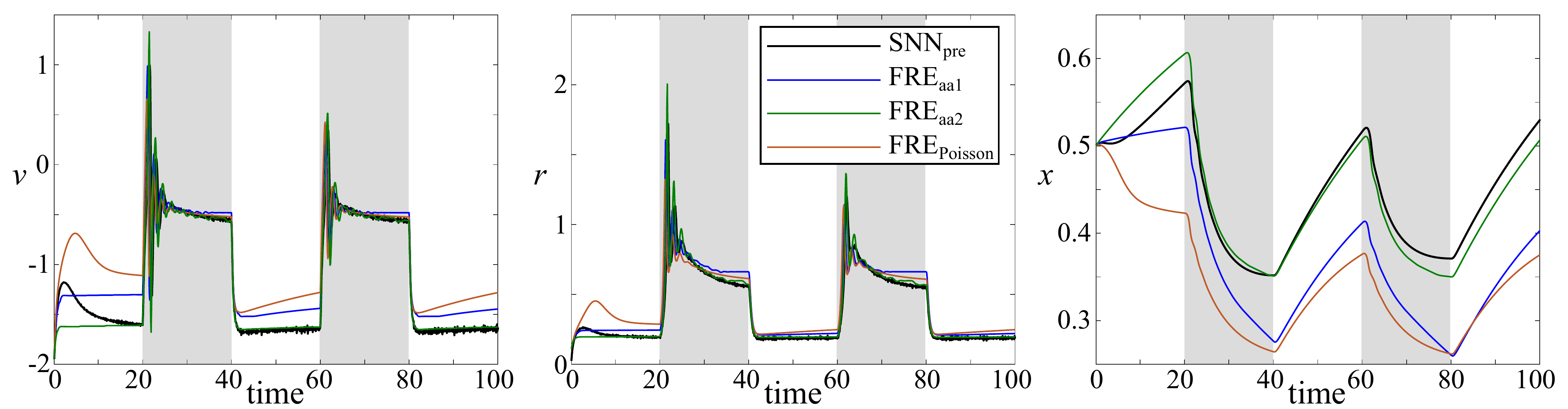}
    \caption{ Comparison of the mean field variables of the microscopic spiking neural network, the mean field model using the Poissonian assumption, and the mean field model with approximation of the effective firing rate. Grey shaded areas indicate time intervals with $I(t) = 3.0$. Parameters: $\alpha = 0.1$, $\tau = 1.0$, $\Delta = 1.0$, $\bar\eta = -2.0$, $J = 15.0$, $\tau_x = 50.0$, $\tau_u = 20.0$, $N = 10000$.}
    \label{fig:approx1}
\end{figure*}

The integrals involved in this approximation are hard to evaluate analytically, 
therefore we solve these integrals numerically for a range of values of $I_r$ and $I_x$ and create look-up tables for $I_r$, $I_x$ and $r_{\mathrm{eff}}$ in order to be able to integrate the resulting model equations numerically.
In Figure~\ref{fig:approx1} we compare the results of the mean-field model $\mathrm{FRE}_{\mathrm{aa}}$ with the dynamics of the spiking neural network $\mathrm{SNN}_{\mathrm{pre}}$, and the mean field model $\mathrm{FRE}_{\mathrm{Poisson}}$.
We find that $\mathrm{FRE}_{\mathrm{aa}}$ is closer to the microscopic dynamics of $\mathrm{SNN}_{\mathrm{pre}}$ than $\mathrm{FRE}_{\mathrm{Poisson}}$.

\section{Conclusion}

In this work, we examined whether spiking neural networks with pre-synaptic short-term plasticity allow for the derivation of low-dimensional mean-field equations via the Lorentzian ansatz described in \cite{montbrio_macroscopic_2015}.
To this end, we considered heterogeneous, all-to-all coupled QIF networks with pre-synaptic STP dynamics, described by a well-known phenomenological model of synaptic short-term depression and facilitation \cite{tsodyks_neural_1998}.
For such QIF networks, other forms of STP have already been shown to be compatible with the Lorentzian ansatz \cite{gast_mean-field_2020}.
In the case of pre-synaptic STP, we identified the evaluation of the effective network input $r_{\mathrm{eff}}$ as the central problem for a mean-field derivation via the Lorentzian ansatz.
This effective network input represents a weighted sum of incoming spikes, where the weights are given by the pre-synaptic depression and facilitation terms.
We presented three different approaches to express $r_{\mathrm{eff}}$ and thus find the mean-field equations:
First, a mean-field description of the STP dynamics via the Poissonian assumption used in \cite{taher_exact_2020};
second, a multi-population approximation that approximates distributed parameters inside the QIF network via a set of coupled sub-populations with different parametrizations;
and third, an adiabatic approximation of the STP time scales.

For the first approach, the effective network input $r_{eff}$ is approximated by a modulation of the mean-field firing rate with an average depression and an average facilitation.
Our analysis revealed that this approach essentially approximates pre-synaptic STP with post-synaptic STP.
We compared the behavior of QIF networks with pre- vs. post-synaptic STP and found that they can express substantial qualitative differences in their dynamics, especially when $\mathrm{SNN}_{\mathrm{pre}}$ expresses a high firing rate heterogeneity across neurons.
Near such regimes, $\mathrm{FRE}_{\mathrm{Poisson}}$ follows the dynamics of $\mathrm{SNN}_{\mathrm{post}}$, and thus fails to capture the behavior of $\mathrm{SNN}_{\mathrm{pre}}$.
It is worth noticing that the mean-field derivation via the Poissonian assumption works well for networks of homogeneous Poisson neurons with independent noise \cite{schmutz_mesoscopic_2020}.
In such networks, single cell firing rates can differ momentarily due to noise, but approach the same rate when averaged over increasing time intervals.
This is a very different scenario compared to the QIF network considered here, where the Lorentzian distribution over $\eta_i$ causes substantial heterogeneity in the single cell firing rates.
Hence, the Poissonian approximation becomes worse the stronger the heterogeneity of single cell firing rates inside the QIF network is. 
In \cite{taher_exact_2020}, where the Poissonian approximation was first applied to a QIF network with pre-synaptic STP, the authors chose QIF networks with relatively low firing rate heterogeneity, leading to a good correspondence with the mean-field model.
Here, we clarified that this correspondence does not generalize to regimes where the QIF network expresses more heterogeneous firing rates.

Populations of neurons that naturally express heterogeneous firing rates exist in sub-cortical structures, for example.
Single cell firing rates in the globus pallidus have been shown to differ substantially across neurons \cite{kita_role_2004,mercer_nav1.6_2007}.
This firing rate heterogeneity has been suggested as an important de-synchronization mechanism of pallidal activity \cite{wilson_active_2013,gast_role_2021}.
Our results suggest that studying the mean-field dynamics in such a population via $\mathrm{FRE}_{\mathrm{Poisson}}$ comes at the risk of substantial errors.
We thus developed a mean-field model that addresses the issue of high firing rate heterogeneities.
Since the distribution over $\eta_i$ is the source of heterogeneity in the QIF network, we attempted to improve the mean-field model by considering a set of coupled sub-networks with distinct, but narrow distributions over $\eta_i$.
This way, the neurons inside each sub-population are parametrized such that they express a considerably lower firing rate heterogeneity than the overall network.
We found that, by increasing the number of sub-populations, the mean-field model converges to the QIF network behavior.
Of course, this approach leads to mean-field models of relatively high dimensionality.
Still, we found that a mean-field model with 100 sub-populations (i.e. a 400-dimensional model), accurately predicted phase transitions of the QIF network from steady-state to oscillatory behavior in a regime where $\mathrm{FRE}_{\mathrm{Poisson}}$ failed to do so.
Thus, we argue that this multi-population approximation provides a flexible mean-field description, the dimensionality of which can be chosen based on the expected firing rate heterogeneity in the neural population under investigation. 

As an alternative to the Poissonian approximation, we applied an adiabatic approximation to the QIF network, assuming slow STP dynamics in comparison to the QIF dynamics.
This assumption is supported by experimental results that suggest depression and facilitation recovery time scales that are at least 10 times slower than typical membrane potential time scales \cite{tsodyks_neural_1997,tsodyks_neural_1998,taher_exact_2020}.
Previously, this approach has been used successfully for the derivation of mean-field equations for QIF networks with spike-frequency adaptation \cite{gast_mean-field_2020}.
By approximating the pre-synaptic STP dynamics as slow, they can be considered as constant, distributed quantities in the fast sub-system.
This way, the STP dynamics do not have to be considered for the evaluation of $r_{\mathrm{eff}}$.
Instead, appropriate distributions over the STP constants have to be chosen.
In our work, we derived analytical solutions of the microscopic STP dynamics in the stationary case and used these solutions to approximate the STP distributions. 
This approach can be considered exact for the description of steady-state solutions, but not for transient dynamics.
That is, the network must have converged to an equilibrium for our approximation to be accurate.
Still, we find that our adiabatic approximation provides a more accurate approximation of the mean-field dynamics of the QIF network dynamics than the Poissonian approximation, even for transient dynamics.
A disadvantage of this method is, however, that we had to approximate the integrals over the STP distribution numerically and calculate $r_{\mathrm{eff}}$ via look-up tables. 
This makes it more difficult to implement the model equations and perform parameter continuations.

In conclusion, we performed a thorough analysis of the problems that arise when attempting to derive the mean-field equations for QIF networks with synaptic short-term plasticity.
Though we did not find a set of exact, closed-form mean-field equations, we provided two different mean-field approximations that we found to be more accurate than a previously proposed mean-field model.
Both of these mean-field approximations can capture the qualitative dynamics of the QIF network and can thus be used for future investigations of its macroscopic dynamics.
Finally, our work provides insight into the distinct effects that pre- vs post-synaptic STP can have on the mean-field dynamics of spiking neural networks.

\section*{Acknowledgements}

R.G. was funded by the Studienstiftung des deutschen Volkes. 
H.S. was supported by the German Research Foundation (DFG (KN 588/7-1) awarded to T.R.K. via Priority Program 2041, “Computational Connectomics”).

\bibliography{apssamp}

\end{document}